\documentclass[11pt]{article}

\usepackage{nicefrac}
\usepackage{fancyhdr}

\begin{document}

\textheight23cm
\topmargin-0.5cm
\textwidth16cm

\newcommand{\troisj}[3]{\left(\begin{array}{ccc}#1 & #2 & #3 \\ 0 & 0 & 0 \end{array}\right)}
\newcommand{\troisjm}[6]{\left(\begin{array}{ccc}#1 & #2 & #3 \\ #4 & #5 & #6 \end{array}\right)}
\newcommand{\sixj}[6]{\left\{\begin{array}{ccc}#1 & #2 & #3 \\ #4 & #5 & #6 \end{array}\right\}}
\newcommand{\neufj}[9]{\left\{\begin{array}{ccc}#1 & #2 & #3 \\ #4 & #5 & #6 \\ #7 & #8 & #9 \end{array}\right\}}

\huge
\begin{center}
A note on recursive calculations of particular $9j$ coefficients
\end{center}

\normalsize

\vspace{0.5cm}

\Large

\begin{center}
Jean-Christophe Pain
\end{center}

\normalsize

\begin{center}
CEA, DAM, DIF, F-91297 Arpajon, France

jean-christophe.pain@cea.fr
\end{center}

\begin{abstract}
The calculation of angular-momentum coupling transformation matrices can be very time consuming and alternative methods, even if they apply only in special cases, are helpful. We present a recursion relation for the calculation of particular $9j$ symbols used in the quantum theory of angular momentum.
\end{abstract} 

\normalsize

\vspace{0.5cm}



\vspace{1cm}


The coupling of $N$ angular momenta is related to the definition of a $3(N-1)j$ symbol. The $9j$ symbols, characterizing the coupling of four angular momenta, are involved for instance in the computation of the matrix elements of the products of tensor operators, and in the transformation from LS to $jj$ coupling \cite{RUDZIKAS04,JUDD63}. The triple sum series of Jucys and Bandzaitis \cite{JUCYS77} is the simplest known algebraic form for the $9j$ coefficient. Fourty years ago, Ali\v{s}auskas and Jucys \cite{ALISAUSKAS71,WU73,BIEDENHARN81,ZHAO88,RAO89} derived an algebraic expression, in which $9j$ symbols are written as the threefold summation of multiplications and divisions of factorials. More recently, Wei \cite{WEI98,WEI99} proposed to express $9j$ symbols as a summation of the products of binomial coefficients and devised an algorithm to calculate the binomial coefficients recursively. Although some relationships between particular $9j$ symbols have been obtained in special cases \cite{GAIGALAS01}, it is quite difficult to find recursion formulas for the $9j$ symbol itself, due to the fact that it contains nine arguments. 

The general recursion relations for arbitrary $9j$ symbols were introduced in \cite{KAROSIENE65}, \cite{JUCYS77} (with incorrect phase factors) and in \cite{VARSHALOVICH88} (in corrected form). However, as a rule in these relations, several angular momentum parameters are changing. It seems that relations with changing single momentum parameter are possible only if some other parameters of the $9j$ accept extreme or fixed small values. For instance, the following $9j$ symbol:

\begin{equation}\label{eq0}
\neufj{\ell_1}{\ell_2}{L}{j_1}{j_2}{L}{\nicefrac{1}{2}}{\nicefrac{1}{2}}{1}
\end{equation}

is often encountered in atomic physics and plays a major role in nuclear physics, especially for the study of $\beta-$decay \cite{deSHALIT63,SZYBISZ75}. It can be evaluated in terms of $3j$ symbols \cite{VARSHALOVICH88}:

\begin{equation}
\neufj{\ell_1}{\ell_2}{L}{j_1}{j_2}{L}{\nicefrac{1}{2}}{\nicefrac{1}{2}}{1}=\frac{1}{\sqrt{6(2L+1)(2j_1+1)(2j_2+1)}}\frac{\troisjm{\ell_1}{\ell_2}{L}{\nicefrac{1}{2}}{\nicefrac{1}{2}}{-1}}{\troisj{L}{j_1}{j_2}},
\end{equation}

but the latter relation is verified only if $j_1$, $j_2$ and $L$ are integers so that $j_1+j_2+L$ is even \cite{VARSHALOVICH88,BRINK68}. More generally, algebraic expressions for $9j$ symbols

\begin{equation}
\neufj{\ell_1}{\ell_2}{L}{j_1}{j_2}{L'}{\nicefrac{1}{2}}{\nicefrac{1}{2}}{1}
\end{equation}

are tabulated as twelve entries of table 10.3 of Ref. \cite{VARSHALOVICH88}, for

\begin{equation}
\neufj{a+\lambda}{b+\mu}{c+\nu}{a}{b}{c}{\nicefrac{1}{2}}{\nicefrac{1}{2}}{1}.
\end{equation}

Moreover, in Appendix 5 of Ref. \cite{JUCYS77}, only two entries for

\begin{equation}
X(b,d,f)=\neufj{b+p}{d+q}{f+r}{b}{d}{f}{\beta}{\delta}{\phi}
\end{equation}

with $\beta=\delta=\nicefrac{1}{2}$, $\phi=1$, $p=q=\nicefrac{1}{2}$ and, respectively, $r=0$ or $r=1$ are included, when the remaining ten entries of table 10.3 of Ref. \cite{VARSHALOVICH88} with $p=\pm\nicefrac{1}{2},q=\pm\nicefrac{1}{2},r=0,\pm 1$ can be generated using some compositions of the ``mirror reflection'' substitutions $b\rightarrow \bar{b}=-b-1,d\rightarrow\bar{d}=-d-1,f\rightarrow\bar{f}=-f-1$ (see Eqs. ($\Pi$.5.2) to ($\Pi$.5.8) in Appendix 5 of Ref. \cite{JUCYS77}). 

Special $9j$ coefficients under consideration include three or four stretched triplets of angular momentum parameters \cite{SHARP67}. In particular, eight entries of table 10.3 of Ref. \cite{VARSHALOVICH88} with $L'=L\pm 1$ are expressed without sum by means of the following expression \cite{JUCYS77,VARSHALOVICH88}:

\begin{eqnarray}
\neufj{a}{b}{a+b}{d}{e}{f}{g}{h}{a+b+f}&=&(-1)^{a+d-g}\frac{\Delta(a+b+fgh)}{\Delta(adg)\Delta(beh)\Delta(def)}\nonumber\\
& &\times\left[\frac{(2a)!(2b)!(2f)!}{(2a+2b+1)(2a+2b+2f+1)!}\right]^{1/2}\nonumber\\
& &\times\frac{(a+b+g+h+f+1)!(g-a+d)!}{(g+h-a-b-f)!(a+g+d+1)!}\nonumber\\
& &\times\frac{(e-b+h)!(d+e-f)!}{(b+e+h+1)!(d+e+f+1)!},\nonumber\\
& &
\end{eqnarray}

where

\begin{equation}
\Delta(abc)=\left[\frac{(a+b-c)!(a-b+c)!(-a+b+c)!}{(a+b+c+1)!}\right]^{1/2},
\end{equation}

together with different symmetry properties of $9j$ symbols. The following useful relation was mentioned by Jang \cite{JANG68}:

\begin{eqnarray}\label{eqd}
A_{j_1,j_2}(L)\neufj{\ell_1}{\ell_2}{L}{j_1}{j_2}{L+1}{\nicefrac{1}{2}}{\nicefrac{1}{2}}{1}&=&B_{\ell_1,\ell_2,j_1,j_2}(L)\neufj{\ell_1}{\ell_2}{L}{j_1}{j_2}{L}{\nicefrac{1}{2}}{\nicefrac{1}{2}}{1}\nonumber\\
& &+C_{j_1,j_2}(L)\neufj{\ell_1}{\ell_2}{L}{j_1}{j_2}{L-1}{\nicefrac{1}{2}}{\nicefrac{1}{2}}{1},\nonumber\\
& &
\end{eqnarray}

with $S=j_1+j_2+L$, and where

\begin{equation}
A_{j_1,j_2}(L)=[L(2L+3)(S+2)(S-2j_1+1)(S-2j_2+1)(S-2L)]^{\nicefrac{1}{2}},
\end{equation}

\begin{equation}
B_{\ell_1,\ell_2,j_1,j_2}(L)=(2L+1)\left[\bar{j}_1-\bar{j}_2-\bar{L}+\frac{2\bar{L}(\bar{j}_2+\nicefrac{3}{4}-\bar{\ell}_2)}{\bar{\ell}_1+\bar{j}_2-\bar{j}_1-\bar{\ell}_2}\right]
\end{equation}

and

\begin{equation}
C_{j_1,j_2}(L)=[(L+1)(2L-1)(S+1)(S-2j_1)(S-2j_2)(S-2L+1)]^{\nicefrac{1}{2}}.
\end{equation}

We follow the convention of Biedenharn \emph{et al.} \cite{BIEDENHARN52}, $\bar{x}=x(x+1)$. Formula (\ref{eqd}) is a particular case of the general recursion relations for $9j$ symbols \cite{JUCYS77,KAROSIENE65,VARSHALOVICH88}. Each $9j$ symbol depends on $L$ via two of its arguments. Defining for instance $u_L=\neufj{\ell_1}{\ell_2}{L}{j_1}{j_2}{L+1}{\nicefrac{1}{2}}{\nicefrac{1}{2}}{1}$, we have $u_{L-1}=\neufj{\ell_1}{\ell_2}{L-1}{j_1}{j_2}{L}{\nicefrac{1}{2}}{\nicefrac{1}{2}}{1}$, which does not appear in Eq. (\ref{eqd}), except if $(j_1,j_2)$ and $(\ell_1,\ell_2)$ are interchanged. However, one has

\begin{eqnarray}\label{eqj}
\neufj{\ell_1}{\ell_2}{L}{j_1}{j_2}{L+1}{\nicefrac{1}{2}}{\nicefrac{1}{2}}{1}&=&\left[\frac{2L+1}{L+(\ell_1-j_1)(2\ell_1+1)+(\ell_2-j_2)(2\ell_2+1)}-1\right]\nonumber\\
& &\times\left[\frac{L(2L-1)(S-2L)}{(L+1)(2L+3)(S-2L+1)}\right]^{1/2}\nonumber\\
& &\times\left[\frac{(S-2j_1+1)(S-2j_2+1)(S+2)}{(S-2j_1)(S-2j_2)(S+1)}\right]^{1/2}\nonumber\\
& &\times\neufj{\ell_1}{\ell_2}{L}{j_1}{j_2}{L-1}{\nicefrac{1}{2}}{\nicefrac{1}{2}}{1},\nonumber\\
& &
\end{eqnarray}

which can be combined to Eq. (\ref{eqd}) to obtain a relation between the coefficients $\neufj{\ell_1}{\ell_2}{L}{j_1}{j_2}{L+1}{\nicefrac{1}{2}}{\nicefrac{1}{2}}{1}$ and $\neufj{\ell_1}{\ell_2}{L}{j_1}{j_2}{L}{\nicefrac{1}{2}}{\nicefrac{1}{2}}{1}$.

Using the Regge symmetry of Wigner $3j$ symbols \cite{REGGE58}, for $\ell_1-j_1=\ell_2-j_2=\pm\nicefrac{1}{2}$ (separately for both $+$ and $-$ signs) \cite{JUCYS77,VARSHALOVICH88}, one finds, concerning the two other $9j$ symbols involved in Eq. (\ref{eqd}), that in some particular cases, as for the $9j$ symbol of Eq. (\ref{eq0}), explicit formulas in terms of $3j$ symbols do exist \cite{BRINK68,NOMURA89}:

\begin{eqnarray}\label{eq1}
\neufj{\ell_1}{\ell_2}{L}{j_1}{j_2}{L-1}{\nicefrac{1}{2}}{\nicefrac{1}{2}}{1}&=&\frac{(j_1-\ell_1)(2\ell_1+1)+(j_2-\ell_2)(2\ell_2+1)-L}{\sqrt{6L(2L-1)(2L+1)(2j_1+1)(2j_2+1)}}\times\nonumber\\
& &(-1)^{\ell_2+j_2+1/2}\frac{\troisjm{\ell_1}{\ell_2}{L}{\nicefrac{1}{2}}{-\nicefrac{1}{2}}{0}}{\troisj{L-1}{j_1}{j_2}}
\end{eqnarray}

and

\begin{eqnarray}
\neufj{\ell_1}{\ell_2}{L}{j_1}{j_2}{L+1}{\nicefrac{1}{2}}{\nicefrac{1}{2}}{1}&=&\frac{(j_1-\ell_1)(2\ell_1+1)+(j_2-\ell_2)(2\ell_2+1)+L+1}{\sqrt{6(L+1)(2L+1)(2L+3)(2j_1+1)(2j_2+1)}}\times\nonumber\\
& &(-1)^{\ell_2+j_2+1/2}\frac{\troisjm{\ell_1}{\ell_2}{L}{\nicefrac{1}{2}}{-\nicefrac{1}{2}}{0}}{\troisj{L+1}{j_1}{j_2}},
\end{eqnarray}

but in that case $j_1+j_2+L$ must be odd. In this way, six entries from above mentioned twelve entries of table 10.3 of Ref. \cite{VARSHALOVICH88} are covered with three expressions.
Although different approaches are needed for the remaining six entries of table 10.3 of Ref. \cite{VARSHALOVICH88} with $\ell_1-j_1=j_2-\ell_2$, which corresponds to the odd linear combinations of $j_1+j_2+L'$, all twelve tabulated expressions are ratios of the factorized elementary linear functions under the square root. Therefore, it is possible to write the binary recursion relations, for example, which allow to express

\begin{equation}
\neufj{\ell_1}{\ell_2}{L+1}{j_1}{j_2}{L'+1}{\nicefrac{1}{2}}{\nicefrac{1}{2}}{1}
\end{equation}

in terms of

\begin{equation}
\neufj{\ell_1}{\ell_2}{L}{j_1}{j_2}{L'}{\nicefrac{1}{2}}{\nicefrac{1}{2}}{1}
\end{equation}

separately for $L'-L=0,\pm 1$. In order to investigate recursion relations for these $9j$ symbols taken independently, let us start from the following identity \cite{VARSHALOVICH88,SZYBISZ75,JANG68}:

\begin{equation}
\neufj{\ell_1}{\ell_2}{L}{j_1}{j_2}{L}{\nicefrac{1}{2}}{\nicefrac{1}{2}}{1}=\frac{\bar{\ell_1}+\bar{j_2}-\bar{j_1}-\bar{\ell_2}}{\sqrt{3\bar{L}}}\neufj{\ell_1}{\ell_2}{L}{j_1}{j_2}{L}{\nicefrac{1}{2}}{\nicefrac{1}{2}}{0},
\end{equation}

which, as mentioned in Ref. \cite{JANG68}, was quoted incorrectly in Refs. \cite{deSHALIT63,ROTENBERG59}. Using the expression of the $9j$ symbol in term of a $6j$ symbol \cite{ROTENBERG59}:

\begin{equation}
\neufj{\ell_1}{\ell_2}{L}{j_1}{j_2}{L}{\nicefrac{1}{2}}{\nicefrac{1}{2}}{0}=\frac{(-1)^{j_1+\ell_2+L+\nicefrac{1}{2}}}{\sqrt{2(2L+1)}}\sixj{\ell_1}{\ell_2}{L}{j_2}{j_1}{\nicefrac{1}{2}},
\end{equation}

we obtain \cite{VARSHALOVICH88}:

\begin{equation}
\neufj{\ell_1}{\ell_2}{L}{j_1}{j_2}{L}{\nicefrac{1}{2}}{\nicefrac{1}{2}}{1}=\frac{(-1)^{j_1+\ell_2+L+\nicefrac{1}{2}}}{\sqrt{6\bar{L}(2L+1)}}\left(\bar{\ell_1}+\bar{j_2}-\bar{j_1}-\bar{\ell_2}\right)\sixj{\ell_1}{\ell_2}{L}{j_2}{j_1}{\nicefrac{1}{2}}.
\end{equation}

Defining the reduced coefficient

\begin{equation}
Q_L=\sqrt{2L+1}\neufj{\ell_1}{\ell_2}{L}{j_1}{j_2}{L}{\nicefrac{1}{2}}{\nicefrac{1}{2}}{1}
\end{equation}

and using the recursion relation for $6j$ symbols described in Ref. \cite{SCHULTEN75}, we find

\begin{equation}
D_{\ell_1,\ell_2,j_1,j_2}(L+1)~Q_{L+1}-E_{\ell_1,\ell_2,j_1,j_2}(L)~Q_L+D_{\ell_1,\ell_2,j_1,j_2}(L)~Q_{L-1}=0,
\end{equation}

where

\begin{eqnarray}
D_{\ell_1,\ell_2,j_1,j_2}(L)&=&\left((L^2-1)(L^2-(\ell_1-\ell_2)^2)((\ell_1+\ell_2+1)^2-L^2)\times\right.\nonumber\\
& &\left.(L^2-(j_2-j_1)^2)((j_2+j_1+1)^2-L^2)\right)^{1/2}
\end{eqnarray}

and

\begin{eqnarray}
E_{\ell_1,\ell_2,j_1,j_2}(L)&=&(2L+1)\left(\bar{L}(-\bar{L}+\ell_1(\ell_1+1)+\ell_2(\ell_2+1))\right.\nonumber\\
& &+j_2(j_2+1)(\bar{L}+\ell_1(\ell_1+1)-\ell_2(\ell_2+1))\nonumber\\
& &+j_1(j_1+1)(\bar{L}-\ell_1(\ell_1+1)+\ell_2(\ell_2+1))\nonumber\\
& &\left.-3\bar{L}/2\right).
\end{eqnarray}

It can be better, in case of large arguments in the $9j$ symbols, to work with the ratio of consecutive elements \cite{LUSCOMBE98}:

\begin{equation}
R_L=\frac{Q_L}{Q_{L-1}},
\end{equation}

which obeys the following recursion relation:

\begin{equation}
R_{L+1}=F_{\ell_1,\ell_2,j_1,j_2}(L)~R_L+G_{\ell_1,\ell_2,j_1,j_2}(L)
\end{equation}

with 

\begin{equation}
F_{\ell_1,\ell_2,j_1,j_2}(L)=\frac{E_{\ell_1,\ell_2,j_1,j_2}(L)}{D_{\ell_1,\ell_2,j_1,j_2}(L+1)}
\end{equation}
 
and 

\begin{equation} 
G_{\ell_1,\ell_2,j_1,j_2}(L)=-\frac{D_{\ell_1,\ell_2,j_1,j_2}(L)}{D_{\ell_1,\ell_2,j_1,j_2}(L+1)}.
\end{equation}

The $9j$ symbol of Eq. (\ref{eq1}) can be formulated in terms of two $6j$ symbols \cite{VARSHALOVICH88}:

\begin{eqnarray}
& &\neufj{\ell_1}{\ell_2}{L}{j_1}{j_2}{L-1}{\nicefrac{1}{2}}{\nicefrac{1}{2}}{1}(-1)^{j_1+\ell_2+L-\nicefrac{1}{2}}\sqrt{6L(4L^2-1)}=\nonumber\\
& &\sqrt{(L^2-(j_1-j_2)^2)((j_1+j_2+1)^2-L^2)}\sixj{\ell_1}{\ell_2}{L}{j_2}{j_1}{\nicefrac{1}{2}}\nonumber\\
& &+\sqrt{(L^2-(\ell_1-\ell_2)^2)((\ell_1+\ell_2+1)^2-L^2)}\sixj{\ell_1}{\ell_2}{L-1}{j_2}{j_1}{\nicefrac{1}{2}}.\nonumber\\
& &
\end{eqnarray}

Each of the two $6j$ symbols involved in the latter identity can be evaluated by recurrence, but we did not find any recursion relation for the $9j$ symbol itself.


We proposed a recursive calculation for special cases of $9j$ symbols, encountered for instance in the transformation matrices from LS to $jj$ coupling schemes, or in the evaluation of matrix elements of tensor operators. We hope that the recursion relation will be useful for the enumeration and algebraic manipulation of $9j$ symbols.

\end{document}